\documentstyle[aps,prd,preprint,floats,epsf]{revtex}
\tighten
\newcommand{\etal}{\it et al. \rm}
\newcommand{\rt}{\rightarrow}
\newcommand{\pipi}{\pi^+ \pi^-}
\preprint{\vbox{\hbox{BIHEP-EP1\_98-01\hfill}
                \hbox{UH-511-903-98\hfill}}}

\title{Branching fractions for 
{\boldmath $\psi(2S) \rt \gamma \eta ' $} and {\boldmath $\gamma\eta$}}

\author{
J.~Z.~Bai$^{1}$,   J.~G.~Bian$^{1}$,  
I.~Blum$^{11}$, 
Z.~W.~Chai$^{1}$,  G.~P.~Chen$^{1}$,
H.~F.~Chen$^{10}$, 
J.~Chen$^3$,
J.~C.~Chen$^{1}$,  Y.~Chen$^{1}$,    Y.~B.~Chen$^{1}$,
Y.~Q.~Chen$^{1}$,  B.~S.~Cheng$^{1}$  X.~Z.~Cui$^{1}$,   H.~L.~Ding$^{1}$,
L.~Y.~Ding$^{1}$,  L.~Y.~Dong$^{1}$,  Z.~Z.~Du$^{1}$,    
W.~Dunwoodie$^7$,
S.~Feng$^{1}$, 
C.~S.~Gao$^{1}$,
M.~L.~Gao$^{1}$,   S.~Q.~Gao$^{1}$,   
P.~Gratton$^{11}$,
J.~H.~Gu$^{1}$,    S.~D.~Gu$^{1}$,
W.~X.~Gu$^{1}$,    Y.~F.~Gu$^{1}$,    Y.~N.~Guo$^{1}$,   S.~W.~Han$^{1}$,
Y.~Han$^{1}$,      
F.~A.~Harris$^8$,
J.~He$^{1}$,       J.~T.~He$^{1}$,    M.~He$^{5}$,
D.~G.~Hitlin$^2$,
G.~Y.~Hu$^{1}$,    H.~M.~Hu$^{1}$,    J.~L.~Hu$^{1}$,    Q.~H.~Hu$^{1}$,
T.~Hu$^{1}$,       X.~Q.~Hu$^{1}$,    J.~D.~Huang$^{1}$, Y.~Z.~Huang$^{1}$,
J.~M.~Izen$^{11}$,
C.~H.~Jiang$^{1}$, Y.~Jin$^{1}$,      Z.~J.~Ke$^{1}$,    
M.~H.~Kelsey$^2$,  B.~K.~Kim$^{11}$,  D.~Kong$^8$,
Y.~F.~Lai$^{1}$,
P.~F.~Lang$^{1}$,  
A.~Lankford$^{9}$,
C.~G.~Li$^{1}$,    D.~Li$^{1}$,       H.~B.~Li$^{1}$,
J.~Li$^{1}$,       P.~Q.~Li$^{1}$,    R.~B.~Li$^{1}$,    W.~Li$^{1}$,
W.~D.~Li$^{1}$,    W.~G.~Li$^{1}$,    X.~H.~Li$^{1}$,    X.~N.~Li$^{1}$,
H.~M.~Liu$^{1}$,   J.~Liu$^{1}$,      J.~H.~Liu$^{1}$,   R.~G.~Liu$^{1}$,
Y.~Liu$^{1}$,      
X.~C.~Lou$^{11}$,  B.~Lowery$^{11}$,
F.~Lu$^{1}$,       J.~G.~Lu$^{1}$,    J.~Y.~Lu$^{1}$,
L.~C.~Lu$^{1}$,    C.~H.~Luo$^{1}$,   A.~M.~Ma$^{1}$,    E.~C.~Ma$^{1}$,
J.~M.~Ma$^{1}$,    
R.~Malchow$^3$,
H.~S.~Mao$^{1}$,   Z.~P.~Mao$^{1}$,   X.~C.~Meng$^{1}$,
J.~Nie$^{1}$,      
S.~L.~Olsen$^8$,   J.~Oyang$^2$,      D.~Paluselli$^8$, L.~J.~Pan$^8$, 
J.~Panetta$^2$,    F.~Porter$^2$,
N.~D.~Qi$^{1}$,    X.~R.~Qi$^{1}$,    C.~D.~Qian$^{6}$,
J.~F.~Qiu$^{1}$,   Y.~H.~Qu$^{1}$,    Y.~K.~Que$^{1}$,   G.~Rong$^{1}$,
M.~Schernau$^9$,
Y.~Y.~Shao$^{1}$,  B.~W.~Shen$^{1}$,  D.~L.~Shen$^{1}$,  H.~Shen$^{1}$,
X.~Y.~Shen$^{1}$,  H.~Y.~Sheng$^{1}$, H.~Z.~Shi$^{1}$,   X.~F.~Song$^{1}$,
J.~Standifird$^{11}$,
F.~Sun$^{1}$,      H.~S.~Sun$^{1}$,   S.~Q.~Tang$^{1}$,  
W.~Toki$^3$,
G.~L.~Tong$^{1}$,
F.~Wang$^{1}$,     L.~S.~Wang$^{1}$,  L.~Z.~Wang$^{1}$,  M.~Wang$^{1}$,
Meng~Wang$^{1}$,   P.~Wang$^{1}$,     P.~L.~Wang$^{1}$,  S.~M.~Wang$^{1}$,
T.~J.~Wang$^{1}$\cite{atNU0},  Y.~Y.~Wang$^{1}$,  
M.~Weaver$^2$,
C.~L.~Wei$^{1}$,   Y.~G.~Wu$^{1}$,
D.~M.~Xi$^{1}$,    X.~M.~Xia$^{1}$,   P.~P.~Xie$^{1}$,   Y.~Xie$^{1}$,
Y.~H.~Xie$^{1}$,   W.~J.~Xiong$^{1}$, C.~C.~Xu$^{1}$,    G.~F.~Xu$^{1}$,
S.~T.~Xue$^{1}$,   J.~Yan$^{1}$,      W.~G.~Yan$^{1}$,
C.~M.~Yang$^{1}$,  C.~Y.~Yang$^{1}$,  J.~Yang$^{1}$,     
W.~Yang$^3$,
X.~F.~Yang$^{1}$,
M.~H.~Ye$^{1}$,    S.~W.~Ye$^{10}$,    Y.~X.~Ye$^{10}$,    K.~Yi~$^{1}$,
C.~S.~Yu$^{1}$,    C.~X.~Yu$^{1}$,    Y.~H.~Yu$^{4}$,   Z.~Q.~Yu$^{1}$,
Z.~T.~Yu$^{1}$,    C.~Z.~Yuan$^{1}$,  Y.~Yuan$^{1}$,     B.~Y.~Zhang$^{1}$,
C.~C.~Zhang$^{1}$, D.~H.~Zhang$^{1}$, Dehong ~Zhang$^{1}$,
H.~L.~Zhang$^{1}$, J.~Zhang$^{1}$,    J.~L.~Zhang$^{1}$, J.~W.~Zhang$^{1}$,
L.~S.~Zhang$^{1}$, Q.~J.~Zhang$^{1}$, S.~Q.~Zhang$^{1}$, X.~Y.~Zhang$^{5}$,
Y.~Zhang$^{1}$,    Y.~Y.~Zhang$^{1}$, D.~X.~Zhao$^{1}$,  
H.~W.~Zhao$^{1}$\cite{atNU1},
J.~W.~Zhao$^{1}$,  M.~Zhao$^{1}$,    W.~R.~Zhao$^{1}$, Z.~G.~Zhao$^{1}$,
 J.~P.~Zheng$^{1}$,
L.~S.~Zheng$^{1}$, Z.~P.~Zheng$^{1}$, G.~P.~Zhou$^{1}$,  H.~S.~Zhou$^{1}$,
L.~Zhou$^{1}$,     Q.~M.~Zhu$^{1}$,   Y.~C.~Zhu$^{1}$,   Y.~S.~Zhu$^{1}$,
B.~A.~Zhuang$^{1}$
\\ (BES Collaboration)}

\address{
$^1$Institute of High Energy Physics, Beijing 100039, People's Republic of
 China\\
$^2$California Institute of Technology, Pasadena, California 91125\\
$^3$Colorado State University, Fort Collins, Colorado 80523\\
$^4$Hangzhou University, Hangzhou 310028,
People's Republic of China\\
$^5$Shandong University, Jinan 250100, People's Republic of
 China\\
$^6$Shanghai Jiaotong University, Shanghai 200030,
People's Republic of China\\
$^7$Stanford Linear Accelerator Center, Stanford, California 94309\\
$^8$University of Hawaii, Honolulu, Hawaii 96822\\
$^9$University of California at Irvine, Irvine, California 92717\\
$^{10}$University of Science and Technology of China, Hefei 230026,
People's Republic of China\\
$^{11}$University of Texas at Dallas, Richardson, Texas 75083-0688}

\begin{document}
\maketitle

\begin{abstract}
We report first measurements of the branching fractions:
${\cal B}(\psi_{2S}\rt \gamma\eta') = (1.54\pm 0.31 \pm 0.20)\times 10^{-4}$
 and
${\cal B}(\psi_{2S}\rt \gamma\eta) =(0.53 \pm 0.31 \pm 0.08) \times 10^{-4}$.
The $\psi(2S)\rt\gamma\eta'$ result is 
consistent with expectations of a model that considers
the possibility of $\eta'$-$\eta_c$ mixing.  The ratio of
the $\psi(2S)\rt\gamma\eta'$ and $\psi(2S)\rt\gamma\eta$ rates
is used to determine the pseudoscalar octet-singlet
mixing angle. 

\end{abstract}



\section{Introduction}
\noindent
In contrast to the $J/\psi$, experimental results on radiative 
decays of the $\psi(2S)$ to non-charmonium hadrons are scarce;
in the latest Particle Data Group tables, 
only upper limits for a few decay modes are listed~\cite{PDG}.
Moreover, it is found experimentally that, while decays to $\rho\pi$ and 
$K^*\overline{K}$ Vector Pseudoscalar ({\em VP}) final states are
significant ($\sim1\%$) for the $J/\psi$, hadronic 
decays of the $\psi(2S)$ to these same {\em VP} final states are
strongly suppressed~\cite{rhopi1,rhopi2}.  This longstanding mystery
of charmonium physics is referred to in the literature as the $\rho\pi$
puzzle~\cite{puzzle}.  The processes $J/\psi\rt\gamma\eta'(958)$ and 
$\gamma\eta$ are radiative {\em VP} channels that have been 
measured by several experiments~\cite{PDG}.  It is of interest to see 
if the same radiative {\em VP} decays of the $\psi(2S)$ are 
suppressed to the same extent as the hadronic $\rho\pi$ and 
$K^*\overline{K}$ decays.

Recently, the CLEO experiment has reported an anomalously large 
branching fraction for the inclusive production of $\eta'$ in the 
$B$-meson decay $B\rt\eta' X_s$, where $X_s$ denotes an inclusive 
hadronic system containing a strange quark~\cite{CLEO}.  One possible
interpretation is the presence of an intrinsic charm 
component of the $\eta'$ meson induced by the strong coupling of the 
$\eta'$ to gluons via the QCD axial anomaly~\cite{mix}.  The resulting
$\eta'$-$\eta_c$ mixing has been proposed as the 
dominant mechanism for the OZI forbidden radiative charmonium decays 
such as $\psi(nS)\rt\gamma\eta'$ and $\gamma\eta$.  In this case, the 
branching fraction for $\psi(2S)\rt\gamma\eta'$ is estimated to be
in the range $(1.0~-~2.7)\times 10^{-4}$~\cite{Chao}.

The ratio of
the $\gamma\eta'$ and $\gamma\eta$ decay rates of the $J^{PC}=1^{--}$
charmonium states is sensitive to the pseudoscalar octet-singlet
mixing angle $\theta_p$.  Assuming that the process
occurs primarily through radiation of the photon from one of the
initial state $c$-quarks and the applicability of SU(3) symmetry for
the decay amplitudes,
one has the simple
relation~\cite{Cahn}
\begin{equation}
\frac{\Gamma(\psi_{nS}\rt\gamma\eta')}{\Gamma(\psi_{nS}\rt\gamma\eta)}=
\left(\frac{p_{\eta'}}{p_{\eta}}\right)^3\frac{1}{\tan^2\theta_p},
\end{equation}
where $p_{\eta}$ ($p_{\eta'}$) is the momentum of the $\eta$ ($\eta'$)
in the $\psi(nS)$ rest frame.  The
measured $J/\psi$ branching fraction values~\cite{PDG}
imply a mixing angle of 
$\vert\theta_p\vert = 22^{\circ} \pm 1^{\circ} \pm 4^{\circ},$ 
(the second error is theoretical), which agrees
well with the value determined from other processes~\cite{Gilman}.
Measurements of the
corresponding branching fractions for the $\psi(2S)$ provide a
consistency check of this relation.  

In this report we present the first measurement of the branching fractions
for $\psi(2S)\rt\gamma\eta'$ and $\psi(2S)\rt\gamma\eta$
using $3.7\times 10^6$ $\psi(2S)$ decays collected using the 
Beijing Spectrometer (BES) 
located at the Beijing Electron-Positron Collider
(BEPC) at the Beijing Institute of
High Energy Physics.

\section{The BES detector}
\noindent
The BES is a large solid-angle magnetic spectrometer 
that is described in detail in ref.~\cite{BES}.  Charged
particle momenta are determined with a resolution of
$\sigma_p/p = 1.7\%\sqrt{1+p^2(GeV^2)}$ in a 40-layer cylindrical 
drift chamber.  Radially outside of the drift chamber is a 
12-radiation-length barrel shower counter (BSC) comprised of
gas proportional tubes interleaved with lead sheets.  The BSC
measures the energies and directions of photons with resolutions
of $\sigma_E/E\simeq 22\%/\sqrt{E(GeV)}$, $\sigma_\phi=4.5$~mrad,
and $\sigma_\theta = 12$~mrad. The iron flux return of the
magnet is instrumented with three double layers of counters
that are used to identify muons.    

For this analysis we  use charged tracks 
with momentum greater than 80~MeV/c that are well fit
to a helix originating near the interaction point.
Candidate $\gamma$'s are
associated with energy clusters in the BSC
that have more than three hit tubes in at least two readout layers.
We use charged tracks and $\gamma$'s that are within the polar angle
region $|\cos\theta|<0.8$.   We reject tracks that are identified
as muons, or that produce high energy showers in the BSC that are
characteristic of electrons.  When computing energies, each
charged track is assigned the pion mass. 

\section{\boldmath Determination of ${\cal B}(\psi(2S)\rt\gamma\eta')$}
\noindent
For the $\psi(2S)\rt\gamma\eta'$ measurement
we investigate the decay chains
\setlength{\unitlength}{2mm}
\begin{center}
\begin{picture}(50,9)(0,0)
\put(0,0)
{\begin{picture}(25,8)(0,0)
  \put(1.3,7.5){\makebox(0,0)[lc]{$\psi_{2S}\rt\gamma\,\eta'$}}
  \put(8.5,6){\line(0,-1){2}}
  \put(8.5,4){\vector(1,0){2.5}}
  \put(12,4){\makebox(0,0)[l]{$\gamma\,\rho^o$}}
  \put(13.7,2.5){\line(0,-1){2}}
  \put(13.7,0.5){\vector(1,0){2.5}}
  \put(16.9,0.7){\makebox(0,0)[l]{$\pi^+\pi^-$}}
 \end{picture}
}
\put(15,6.8){and}
\put(21.5,00)
{\begin{picture}(25,8)(0,0)
  \put(0.3,7.5){\makebox(0,0)[lc]{$\psi_{2S}\rt\gamma\,\eta'$}}
  \put(8,6){\line(0,-1){2}}
  \put(8,4){\vector(1,0){2.5}}
  \put(11.3,4){\makebox(0,0)[l]{$\pi^+\pi^-\eta$}}
  \put(16.5,2.5){\line(0,-1){2}}
  \put(16.5,0.5){\vector(1,0){2.5}}
  \put(19.5,0.5){\makebox(0,0)[l]{$2\gamma$.}}
 \end{picture}
}
\end{picture}
\end{center}
\noindent
It follows that the
reactions of interest are $\psi(2S)\rt\pi^+\pi^-\gamma\gamma$ for 
the  $\rho^0\gamma$ mode, and 
$\psi(2S)\rt\pi^+\pi^-\gamma\gamma\gamma$ for the $\pipi\eta$ decays. 

\subsection{\boldmath The $\psi(2S)\rt\gamma\eta'\rt
\gamma\gamma\rho^0$ measurement}
\begin{figure}[!h]
\centerline{\epsfysize 2.0 truein
\epsfbox{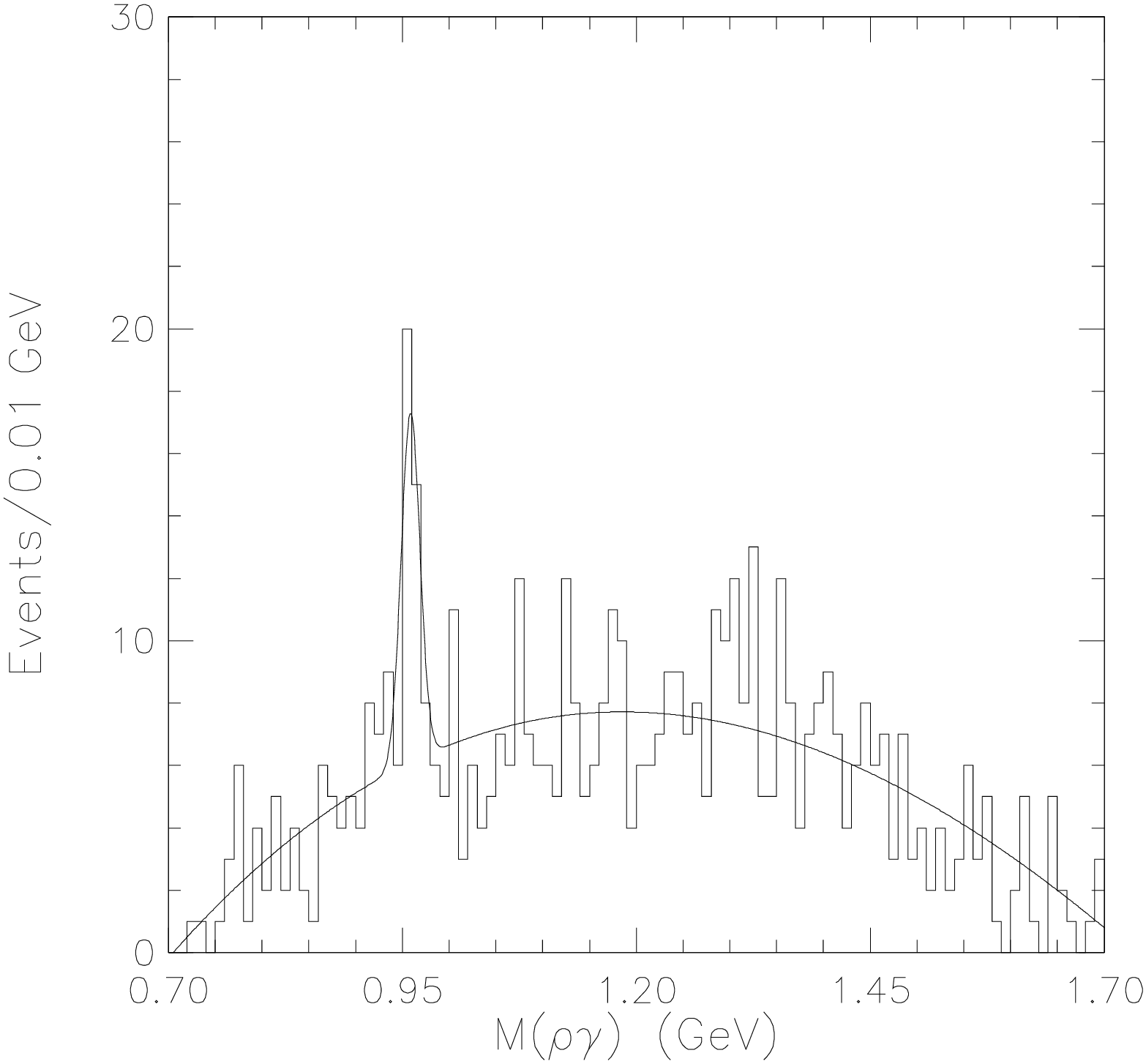}}
\caption{\label{Mpipigam}
The $\pipi\gamma$ invariant mass distribution for selected events.}
\end{figure}

\noindent
For the measurement using the  $\eta'\rt\gamma\rho^0$ mode,
we require two oppositely charged tracks 
with an opening angle $\theta_{open}<130^{\circ}$ and at least
two candidate $\gamma$'s that are more than $10$ degrees 
away from the nearer charged track.
The events where the total energy of the two charged tracks
is less than 2.1~GeV
are subjected to
a four-constraint
kinematic fit to the hypothesis $\psi(2S)\rt\pi^+\pi^-\gamma\gamma$,
and required to
have $\chi^2<15$. The $\pi^+\pi^-\gamma$ mass distribution
for events with  $M_{\pi^+\pi^-}$ within $0.15$~GeV of $M_{\rho}$
and a  $\gamma\gamma$ opening angle greater than $110^{\circ}$
is plotted in Fig.~\ref{Mpipigam},
where a peak at the mass of the $\eta'(958)$ is apparent.

The curve in Fig.~\ref{Mpipigam} is the result of a fit to the 
measured mass distribution with the $\eta(958)$ represented as a
Gaussian, and a
third-order polynomial background 
function.   The
width of the Gaussian is fixed at the Monte-Carlo determined
experimental resolution of $\sigma=0.01$~GeV~\cite{mc_res}. 
The fitted Gaussian has a peak position at $M_{\eta(958)}$ and
$N_{evts}=28.1\pm 7.2$ events.  
Events from the cascade decays $\psi(2S)\rt anything + J/\psi$, where
$J/\psi\rt\gamma\eta^{'}$ or $J/\psi\rt\pi^0\rho^0$,  
also can give a peak at 
$M_{\eta'(958)}$.  We subject a sample of Monte Carlo simulated
events equivalent to
ten times the $\psi(2S)$ data set to the same selection
and fitting procedure. The resulting 
estimate of the contamination from this source is
$N_{bkg}=1.4\pm0.5$ events, where the error is
statistical and comes from the fit.

We use Monte Carlo simulated
events to determine the acceptance.  
The events are generated with a $1 + \cos^2\theta$ angular
distribution for the $\psi(2S)\rt\gamma\eta'$ decays, an isotropic
distribution for the $\eta'\rt\gamma\rho^0$ decays, followed by
helicity $\pm 1$
$\rho^0\rt\pi^+\pi^-$ decays. The 
acceptance determined in this way is $\epsilon_{\rho\gamma}=0.18\pm0.02$,
where the error includes both Monte Carlo statistics (7\%) and uncertainties
in the simulation program (8\%)~\cite{mc_eff} added in quadrature.

The $\psi(2S) \rt \gamma\eta'$  branching fraction is 
determined from the relation
\begin{eqnarray}
\nonumber
{\cal B}(\psi_{2S} \rt \gamma\eta') & = &
\frac{N_{evts} - N_{bkg}}{N_{\psi_{2S}} {\cal B}(\eta' \rt \gamma\rho)
\epsilon_{\rho\gamma}}\\
& = & (1.36 \pm 0.37 \pm 0.20) \times 10^{-4}.
\end{eqnarray}
Here the first error is statistical, and the second 
is the systematic error due to uncertainties
in $N_{\psi_{2S}}$ (9\%), the acceptance (11\%)
and the $\eta' \rt \gamma\rho$ branching fraction (3.3\%)
added in quadrature.

\subsection{\boldmath The $\psi(2S)\rt\gamma\eta'\rt
\gamma\pi^+\pi^-\eta$ measurement}
\begin{figure}[!h]
\centerline{\epsfysize 2.0 truein
\epsfbox{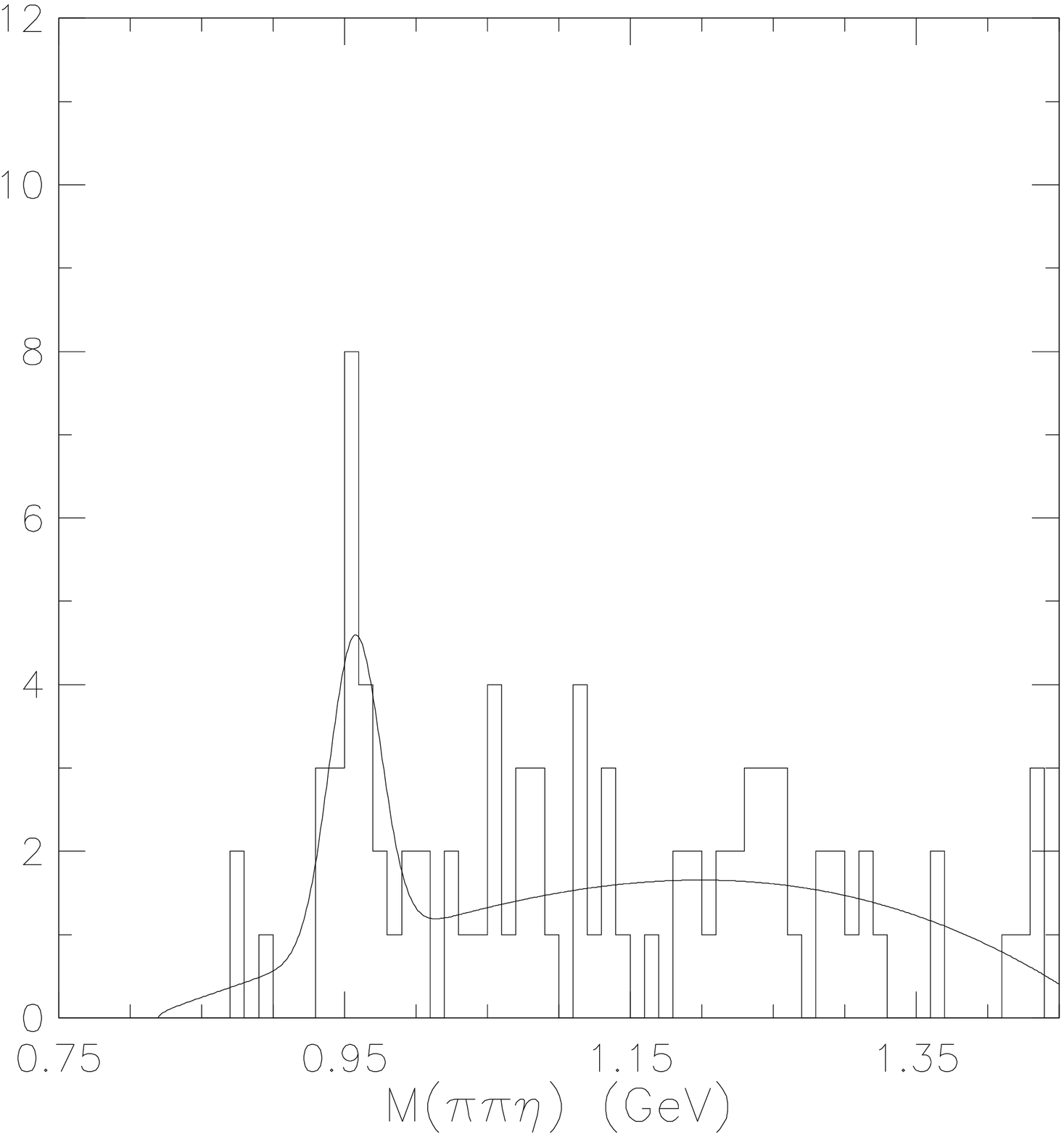}}
\caption{\label{Mpipieta}
The $\pipi\eta$ invariant mass distribution for selected events.}
\end{figure}

\noindent
For the measurement using the  $\eta'\rt\pi^+\pi^-\eta$ mode,
we require two oppositely charged tracks 
with an opening angle $\theta_{open}<70^{\circ}$, and at least
three candidate $\gamma$'s that are more than five degrees 
away from the nearer charged track.  We select
events with a total energy for the two charged tracks
that is less than 1.2~GeV, and 
require them to satisfy
a four-constraint
kinematic fit to the hypothesis $\psi(2S)\rt\pi^+\pi^-\gamma\gamma\gamma$
with $\chi^2<12$.  We identify $\gamma\gamma$ pairs with an invariant
mass within $0.03$~GeV of $M_{\eta}$ as candidate $\eta$'s.
The $\pi^+\pi^-\eta$ mass distribution
for the selected events
is plotted in Fig.~\ref{Mpipieta}.
There is a  peak in the data at the mass of the $\eta'(958)$.

The curve in Fig.~\ref{Mpipieta} is the result of a fit to the 
measured mass distribution with the $\eta(958)$ represented as a
Gaussian, and a polynomial background 
function that is forced to zero at the
$\pi^+\pi^-\eta$ threshold. The width of the Gaussian is fixed
at $\sigma=0.018$~GeV, the resolution value determined from
the MC simulation.
The fitted Gaussian has
$N_{evts}=16.8\pm 4.9$ events.  The Monte Carlo estimate of
backgrounds from $\psi(2S)\rt J/\psi$ cascade decays is
$N_{bkg}=0.35\pm0.03$.
The MC determined acceptance for this mode is $0.14\pm0.015$,
and the corresponding  $\psi(2S) \rt \gamma\eta'$  branching 
fraction is
\begin{equation}
{\cal B}(\psi_{2S} \rt \gamma\eta')  = 
(2.00 \pm 0.59 \pm 0.29) \times 10^{-4}.
\end{equation}
The agreement with the result determined for the $\eta'\rt\gamma\rho^0$
mode is reasonable.

\section{\boldmath The $\psi(2S)\rt\gamma\eta$ measurement}
\noindent
For the $\psi(2S)\rt\gamma\eta$ measurement, we use the
$\eta\rt\pi^+\pi^-\pi^0$ decay mode.  This corresponds to the
same $\psi(2S)\rt\pi^+\pi^-\gamma\gamma\gamma$ reaction
as for the $\eta'\rt\pi^+\pi^-\eta$ measurement.

\begin{figure}[!h]
\centerline{\epsfysize 2.0 truein
\epsfbox{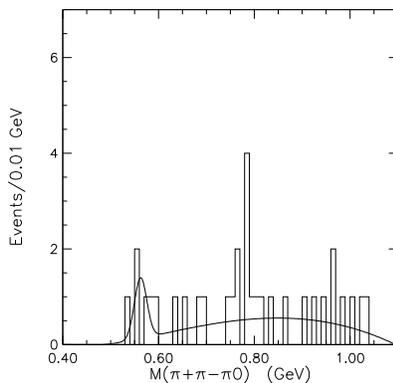}}
\caption{\label{Mpipipi0}
The $\pipi\pi^0$ invariant mass distribution for selected events.
The region near the $\omega(785)$ is excluded from the fit.}
\end{figure}

We require two oppositely charged tracks 
with total energy less than 1.7~GeV, and at least
three candidate $\gamma$'s that are more than five degrees 
away from the nearer charged track.
The events
are required to satisfy
a four-constraint
kinematic fit to the hypothesis $\psi(2S)\rt\pi^+\pi^-\gamma\gamma\gamma$
with $\chi^2<12$.  We identify $\gamma\gamma$ pairs with an invariant
mass within $0.025$~GeV of $m_{\pi^0}$ as candidate $\pi^0$'s.
The $\pi^+\pi^-\pi^0$ mass distribution
is plotted in Fig.~\ref{Mpipipi0}.
Here small clusters of events appear at the mass of the $\eta(547)$
and the $\omega(780).$

The curve in Fig.~\ref{Mpipipi0} is the result of a fit to the 
measured mass distribution with the $\eta$ represented as a
Gaussian, and a polynomial background
function.  The width of the Gaussian is fixed at $\sigma = 0.013$~GeV,
the value determined from the MC simulation.   
The $\omega(780)$ mass region is
excluded from the fit.
The fitted Gaussian has a peak position at $M_{\pipi\pi^0}=0.56\pm0.01$,
which is one standard deviation above $M_{\eta}$, and an area of
$N^{\eta}_{evts}=4.1\pm 2.4$ events. 

As a check, we used the events in the $\pi^0$ sidebands of the $\gamma\gamma$
invariant mass distribution as an experimental estimate of our background.
Here we find no events within $\pm3\sigma$ of $M_{\eta}$ and a fit
to the sideband-subtracted $M_{\pi+\pi-\pi^0}$ distribution
yields $6.0\pm2.5$ $\eta$ events.

The $\psi(2S)\rt\gamma\eta$ signal 
has a statistical significance corresponding to a
little less than $2\sigma$~\cite{signif}.
If we treat the 4.1 observed events as a real signal, the
$\psi(2S) \rt \gamma\eta$  branching fraction is determined to be
\begin{equation}
{\cal B}(\psi_{2S} \rt \gamma\eta)  = 
     (0.53\pm 0.31\pm0.08) \times 10^{-4}.
\end{equation}
(The MC-determined acceptance for this channel is
$0.10\pm0.012$.)
The 4.1 events from the fit imply a 90\% confidence level (c.l.) 
upper limit of 7.2 events; this corresponds to a 90\% 
c.l. limit on the $\psi(2S)\rt\gamma\eta$ branching fraction of 
$0.9\times 10^{-4}.$

\section{Discussion}
\noindent
Combining the two results for
${\cal B}(\psi(2S)\rt \gamma\eta')$ from the different
$\eta'$ decay modes gives~\cite{common}
\begin{equation}
{\cal B}(\psi_{2S}\rt \gamma\eta') = (1.54\pm 0.31 \pm 0.20 )\times 10^{-4},
\end{equation}
which is within the range expected for the case where 
$\eta^{'}-\eta_c$ mixing is important~\cite{Chao}.
To compare with $J/\psi$ decays, we use the ratio
\begin{equation}
Q_{\gamma\eta'} = \frac{{\cal B}(\psi_{2S}\rt \gamma\eta')}
{{\cal B}(J/\psi\rt \gamma\eta')} = 0.036\pm 0.009.
\end{equation}
This low value for $Q_{\gamma\eta'}$
indicates that this $\psi(2S)$ decay mode is suppressed
relative to di-lepton decays, where the corresponding ratio
$Q_{\ell\ell} = 0.147 \pm0.023$~\cite{PDG},
but not as severely as in the case of $\rho\pi$, where
$Q_{\rho\pi}<0.002$, or $K^{*+}K^-$, 
where $Q_{K^{*+}K^-}<0.006$~\cite{rhopi2}.
Pinsky~\cite{Pinsky} relates the processes $\psi(2S)\rt\gamma\eta'$ to 
the hindered $M1$ transition $\psi(2S)\rt\gamma\eta_{c}$.
He predicts $Q_{\gamma\eta'} = 0.002$,
which is well below our measured value.

The suppression of $J/\psi\rt\gamma\eta$  
relative to $J/\psi\rt\gamma\eta'$ decays
appears to also occur for the $\psi(2S)$:
\begin{equation}
\frac{{\cal B}(J/\psi\rt\gamma\eta)}{{\cal B}(J/\psi\rt\gamma\eta')}
= 0.200\pm 0.023 ~({\rm PDG});
\end{equation}
\begin{equation}
\frac{{\cal B}(\psi(2S)\rt\gamma\eta)}{{\cal B}(\psi(2S)\rt\gamma\eta')}
= 0.34\pm 0.22 ~({\rm this~expt}).
\end{equation}
Our results provide an
independent evaluation of the mixing angle of 
$\vert\theta_p\vert = {28^{\circ}}^{+7^{\circ}}_{-10^{\circ}},$
which is consistent with other determinations,
albeit with larger errors.

\section{Acknowledgements}
\noindent
We acknowledge the strong efforts of the BEPC staff and the
helpful assistance we received from the members of the IHEP computing center.
We also thank K-T.~Chao for a number of helpful discussions.


\begin{references}
\bibitem{mc_res} The resolution parameters in the simulation
program have been tuned and checked with
data from high statistics processes.  For example, the 
measured width of the $\sim$400 event $\omega\rt\pi^+\pi^-\pi^0$ peak in
$\psi(2S)\rt\pi^+\pi^-\omega$ decays ($\sigma = 16.1\pm 1.1$~MeV)
agrees well with the MC expectation ($\sigma_{MC}=16.6$~MeV).
\bibitem{mc_eff} The systematic error on the MC efficiency is
mostly due to uncertainties in the modeling of the photon shower
clustering.  These are studied using $J/\psi\rt\pi^+\pi^-\pi^0$
events.
\bibitem{signif} The area under the background function
for the interval $M_{\eta} \pm 3\sigma$ is 0.8 event.  
The probability that this can fluctuate to
the observed signal plus background is $\sim$3.5\%.
\bibitem{common} All components of the systematic error 
calculations for the two $\eta'$ decay modes
are common except for the MC statistics.
\end{references}
\end{document}